# THE CMB DIPOLE:
# THE MOST RECENT MEASUREMENT AND SOME HISTORY

Charles H. Lineweaver

*Université Louis Pasteur*
*Observatoire Astronomique de Strasbourg*
*11 rue de l'Université, 67000 Strasbourg, France*
*charley@cdsxb6.u-strasbg.fr*

## ABSTRACT

The largest anisotropy in the cosmic microwave background (CMB) is the $\approx 3$ mK dipole assumed to be due to our velocity with respect to the CMB. Over the past ten years the precision of our knowledge of the dipole has increased by a factor of ten. We discuss the most recent measurement of this dipole obtained from the four year COBE Differential Microwave Radiometers (DMR) as reported by Lineweaver et al. (1996). The best-fit dipole is $3.358 \pm 0.001 \pm 0.023$ mK in the direction $(\ell, b) = (264°.31 \pm 0°.04 \pm 0°.16, +48°.05 \pm 0°.02 \pm 0°.09)$, where the first uncertainties are statistical and the second include calibration and systematic uncertainties. The inferred velocity of the Local Group is $v_{LG} = 627 \pm 22$ km/s in the direction $\ell = 276° \pm 3$, $b = 30° \pm 2$. We compare this most recent measurement to a compilation of more than 30 years of dipole observations.

# 1 Introduction

The Sun's motion with respect to the rest frame of the cosmic microwave background (CMB) is believed to be responsible for the largest anisotropy seen in the COBE DMR maps: the $\approx 3$ mK dipole in the direction of the constellation Leo. A high precision measurement of this Doppler dipole is important because it

- tells us our velocity with respect to the rest frame of the CMB.

- will be used as the primary calibrator for an increasing number of ground, balloon and satellite anisotropy experiments (Bersanelli et al. 1996). Small scale experiments are becoming sensitive enough to use the dipole to calibrate (Richards 1996). Thus the typical 10-20% absolute calibration accuracy of ground and balloon-borne experiments can be improved by more than an order of magnitude to the 0.7% absolute calibration accuracy of the DMR dipole.

- permits the accurate removal of the Doppler dipole and Doppler quadrupole from anisotropy maps thus improving the precision of the anisotropy results.

- calibrates bulk flow observations which yield independent but much less precise dipole values.

- permits an eventual test of the Doppler origin of the CMB dipole in which it is compared to the dipoles in other background radiations (Lineweaver et al. 1995).

In this paper we discuss the most recent determination of the precise direction and the amplitude of the dipole observed in the DMR four-year data. We discuss contamination from Galactic emission as well as other factors contributing to the error budget (see Lineweaver et al. (1996) for details). We then compare our results to a compilation of more than 30 years of dipole results.

# 2 Minimizing the Error Due to Galactic Foreground, CMB Background and Instrument Noise

## 2.1 Galactic Plane Cuts

We estimate the influence of Galactic emission on the measurement by solving for the dipoles for a series of Galactic plane latitude cuts. The dipole amplitude and direction results from each channel and each Galactic cut are shown in Figure 1. Galactic emission produces a dipole which pulls the solutions towards it. This is easily seen in Figure 1 from the locations of the 0° and 5° cut solutions relative to the cluster of higher cut results on the right. Since the Galactic dipole vector is nearly orthogonal to the CMB dipole vector, it is almost maximally effective in influencing the CMB dipole direction and almost minimally effective in influencing the CMB dipole amplitude.

In Figure 1, the general increase of the dipole amplitudes seen in the top panel as the Galactic cut increases from 0° to 5° to 10° can be explained by the fact that the Galactic dipole vector contains a component in the direction opposite to the CMB dipole (the Galactic center is $\approx 94°$ away) and thus reduces the total dipole in the maps.

Figure 1 clearly shows the influence of the Galaxy for the 0° and 5° cuts as well as the relative agreement of the independent channel results for both amplitude and direction. It is also apparent that to first approximation a $10°$ cut is sufficient to remove the effect of the Galaxy on the direction of the best-fit dipole; increases of the cut from $10°$ to $15°$ and so on, do not push the directions away from the Galactic center or in any other particular direction. The results tend to cluster together. The directional precision of the various channels and Galactic cuts is seen to be $\sim 0°.3$ and it is perhaps reassuring to note that at the bottom and the top of the cluster are the least sensitive 31A and 31B solutions.

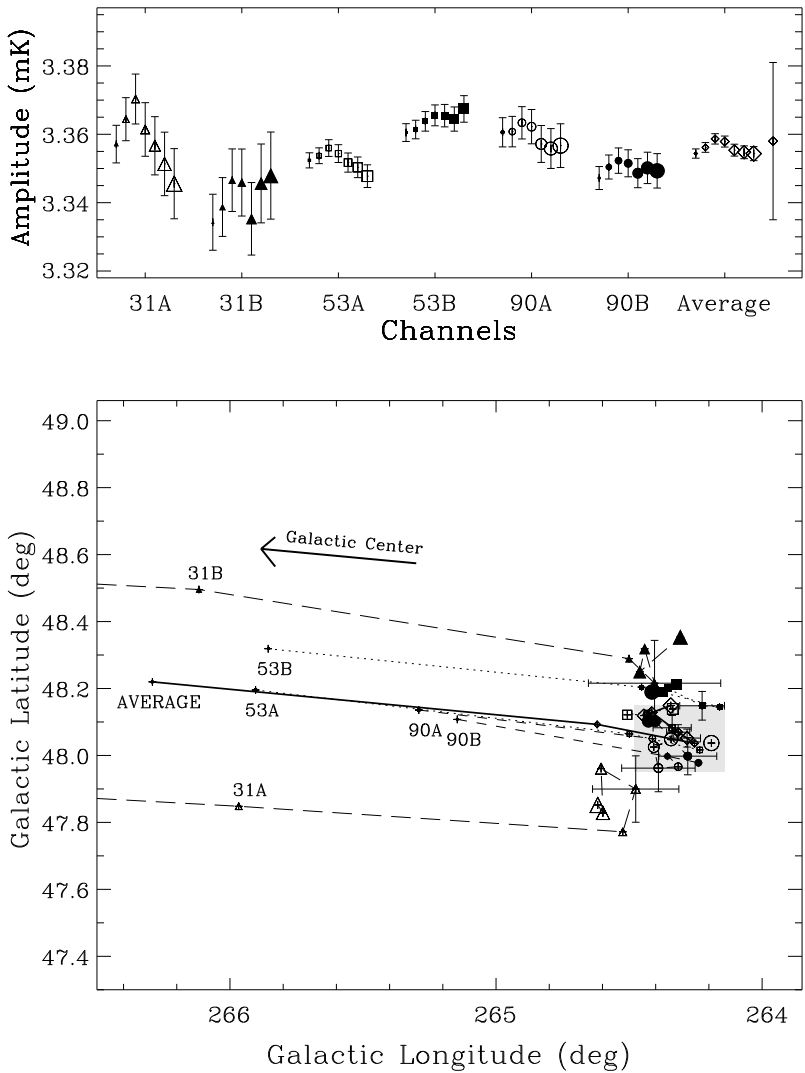

**Figure 1. Dipole Amplitudes (top) and Directions (bottom)** The results for each channel and Galactic plane cut (from left to right in the top panel), $|b| > 0°$, $5°$, $10°$, $15°$, $20°$, $25°$, $30°$ are shown. Channels and cuts are denoted with the same point type and size in both panels. Solutions for the dipole where no effort has been made to eliminate Galactic emission (i.e., $0°$ Galactic cuts) are labeled with the channel names "53A", "53B", "90A" and "90B". The 31 GHz labels indicate the $5°$ cut solutions since their $0°$ cut solutions are off the plot at longitude $\approx 271°$. For each channel, the successive Galactic cuts are connected by lines (31: long dashes, 53: dots, 90: short dashes, Average: solid). The direction of the Galactic center is toward higher latitudes for the same reason that one flies north-west from London to arrive at New York. The latitude and longitude ranges were chosen to display an approximately square piece of the sky. For each channel, the direction error bars on the $15°$ Galactic cut solutions are shown. Our final dipole amplitude, including the calibration uncertainty is the point in the far right of the top panel. The grey box in the bottom panel denotes the 68% confidence levels of our final dipole direction (cf. Fig. 1, Lineweaver et al. (1996)).

Galactic emission significant enough to affect the dipole results will tend to pull the three channels in approximately the same direction and favor a spectral behavior typical of synchrotron or free-free emission. This is easily seen for the $0°$ and $5°$ cuts (cf. Figure 2, Lineweaver et al. 1996). The absence of this behavior for the $10°$ and $15°$ cuts and larger is evidence that the Galaxy is no longer the major contributor to the directional uncertainty of the dipole.

## 2.2 Higher Multipole CMB as Unwanted Contamination

For the purposes of determining the dipole there are two sources of noise; instrument noise with a power law spectral index $n \approx 3$ and the $n \approx 1$ CMB signal. At $10°$ scales the CMB signal to noise ratio in the maps is $\sim 2$ (Bennett et al. 1996). Thus on larger scales the CMB signal dominates the instrument noise and correspondingly, the uncertainties on the dipole from the CMB signal are larger than those from the instrument noise. The uncertainties from *both* are reduced by lowering the Galactic plane cut. This is further supported by the fact that for $|b| \gtrsim 20°$, the combined free-free and dust emission from the Galaxy at 53 and 90 GHz produces only $\sim 10\,\mu K$ rms (Kogut et al. 1996a) while the CMB signal rms is $\sim 35\,\mu K$ (Banday et al. 1996).

To estimate the uncertainty in the dipole results due to the CMB signal we simulate $n = 1.2$, $Q_{rms-PS} = 15.3\,\mu K$ CMB skies for $2 \leq \ell \leq 25$. We superimpose these maps on a known dipole and solve for the dipole using a 15° Galactic plane cut. No bias is detected and the rms's of the results around the input values are $3.3\,\mu K$ in amplitude, $0°.127$ in longitude and $0°.062$ in latitude. We include these CMB contamination uncertainties in our estimate of the systematic errors.

We have found that Galactic cuts greater than 15° are not useful corrections which eliminate more and more Galactic contamination; they introduce systematic errors associated with large Galactic cuts due to the increasingly non-orthogonal basis functions $Y_{\ell m}(\theta, \phi)$, over the increasingly limited and thus noisier input data.

We conclude that the Galactic cuts of 10° and 15° are the best compromise to minimize the combined effect of CMB aliasing, Galactic contamination and noise. The high precision of our dipole direction results depend on this conclusion. Note that this choice for the optimal Galactic cut is smaller than the $\approx 20°$ cut used when one is trying to compute the correlation function or determine the $\ell \geq 2$ components of the power spectrum of the CMB signal which are smaller than the dipole by a factor of $\sim 200$. For such determinations, the similar compromise for simultaneously minimizing Galactic contamination, instrument noise and other procedural/systematic effects demands a larger cut.

## 3 Results

Taking the averages of the 10° and 15° cuts and the weighted average of all six channels we obtain a best-fit dipole amplitude $3.358 \pm 0.001 \pm 0.023$ mK in the direction $(\ell, b) = (264°.31 \pm 0°.04 \pm 0°.16, +48°.05 \pm 0°.02 \pm 0°.09)$, where the first uncertainties are statistical and the second are estimations of the combined systematics. In celestial coordinates the direction is $(\alpha, \delta) = (11^h 11^m 57^s \pm 23^s, -7°.22 \pm 0°.08)$ (J2000). The uncertainty in the dipole amplitude is dominated by the absolute calibration of the DMR instrument (Kogut et al. 1996b). This is easily seen in Figure 1 by comparing the large error bars on our final result (far right) with the noise-only error bars on the channel results. The calibration uncertainty plays no role in the directional uncertainty for the same reason that the directions of vectors $\vec{x}$ and $a\vec{x}$ (where $a$ is any positive constant) are the same.

Under the assumption that the Doppler effect is responsible for the entire CMB dipole, the velocity of the Sun with respect to the rest frame of the CMB is $v_\odot = 369.0 \pm 2.5$ km/s, which corresponds to the dimensionless velocity $\beta = v_\odot/c = 1.231 \pm 0.008 \times 10^{-3}$. The associated rms Doppler quadrupole[1] is $Q_{rms} = 1.23 \pm 0.02\,\mu K$ with components $[Q_1, Q_2, Q_3, Q_4, Q_5] = [0.91 \pm 0.02, -0.20 \pm 0.01, -2.04 \pm 0.03, -0.91 \pm 0.02, 0.18 \pm 0.01]\,\mu K$. The velocity of the Local Group with respect to the CMB can be inferred; following Kogut et al. (1993) we obtain $v_{LG} = 627 \pm 22$ km/s in the direction $\ell = 276 \pm 3$, $b = 30 \pm 2$.

---

[1] $Q_{rms}^2 = \frac{4}{15}[\frac{3}{4}Q_1^2 + Q_2^2 + Q_3^2 + Q_4^2 + Q_5^2]$ where the components are defined by $T_o \frac{\beta^2}{2}(2cos^2\theta - (2/3)) = Q_1(3sin^2 b - 1)/2 + Q_2 sin2b\, cos\ell + Q_3 sin2b\, sin\ell + Q_4 cos^2 b\, cos2\ell + Q_5 cos^2 b\, sin2\ell$, where $T_o$ is the mean CMB temperature and $\theta$ is the angle between the dipole direction and the direction of observation: $(\ell, b)$.

# 4 Historical Discussion

We have compiled more than 30 years of dipole measurements in Table 1 and these numbers were used to make Figure 2. This plot may a good example of scientific progress. We acknowledge support from the French Ministère des Affaires Etrangères.

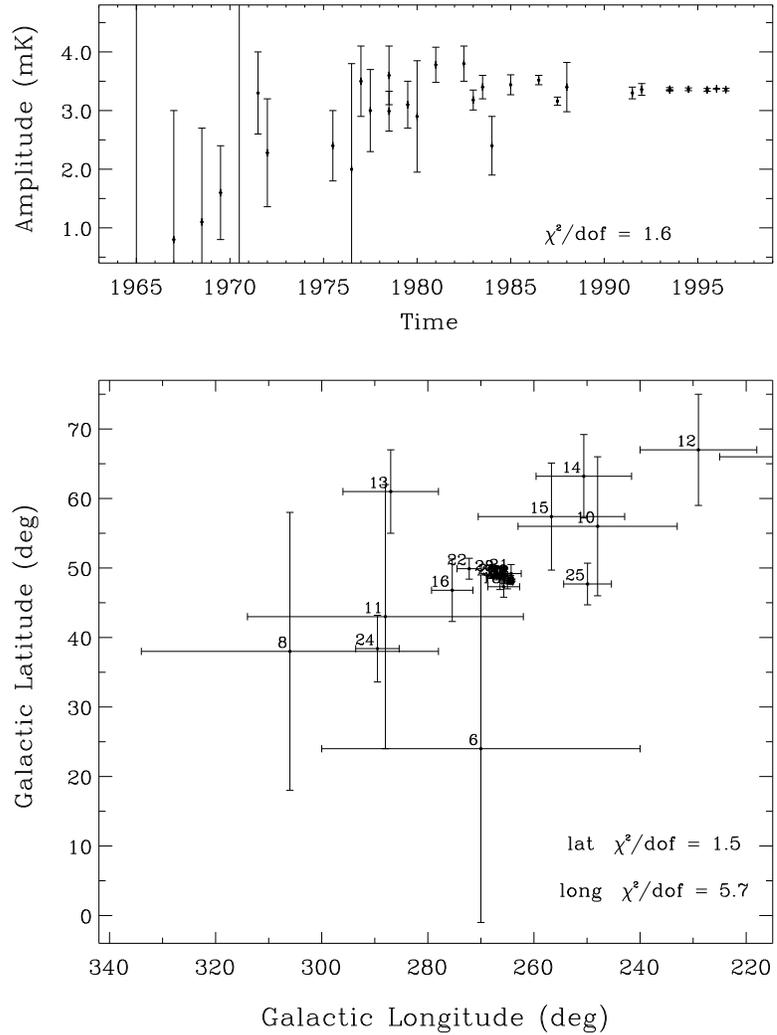

**Figure 2. 30 Years of CMB Dipole Measurements**
These results and the number labels in the bottom panel correspond to the references in Table 1. To see how realistically the dipole community has estimated its errors we have calculated the $\chi^2$ per degree of freedom for the amplitudes, galactic longitudes and latitudes of the reported results. They are respectively 1.6, 5.7 and 1.5 (approximately correct error estimates yield $\chi^2/dof \sim 1$). Thus, the amplitude and latitude estimates are believable while the reported errors on longitude have been underestimated. This can probably be attributed to the various ways in which Galactic emission has (or has not) been accounted for since a line fit to the set of dipole directions passes through the Galactic center ($\ell = 360°, b = 0$). This plot may be a good example of scientific progress.

Table 1: CMB Dipole Measurements

| # | Reference | Amplitude D(mK) | ± σ | Longitude[a] ℓ(deg) | ± σ | Latitude[a] b(deg) | ± σ | Freq (GHz) |
|---|---|---|---|---|---|---|---|---|
| 1 | Penzias & Wilson(1965) | < 270 | | | | | | 4 |
| 2 | Partridge & Wilkinson(1967) | 0.8 | 2.2 | | | | | 9 |
| 3 | Wilkinson & Partridge(1969) | 1.1 | 1.6 | | | | | 9 |
| 4 | Conklin(1969) | 1.6 | 0.8 | 96 | 30 | 85 | 30 | 8 |
| 5 | Boughn et al. (1971) | 7.6 | 11.6 | | | | | 37 |
| 6 | Henry(1971) | 3.3 | 0.7 | 270 | 30 | 24 | 25 | 10 |
| 7 | Conklin(1972) | > 2.28 | 0.92 | 195 | 30 | 66 | 10 | 8 |
| 8 | Corey & Wilkinson(1976) | 2.4 | 0.6 | 306 | 28 | 38 | 20 | 19 |
| 9 | Muehler(1976) | 2.0 | 1.8 | 207 | | −11 | | 150 |
| 10 | Smoot et al. (1977) | 3.5 | 0.6 | 248 | 15 | 56 | 10 | 33 |
| 11 | Corey(1978) | 3.0 | 0.7 | 288 | 26 | 43 | 19 | 19 |
| 12 | Gorenstein(1978) | 3.60 | 0.5 | 229 | 11 | 67 | 8 | 33 |
| 13 | Cheng et al. (1979) | 2.99 | 0.34 | 287 | 9 | 61 | 6 | 30 |
| 14 | Smoot & Lubin(1979) | 3.1 | 0.4 | 250.6 | 9 | 63.2 | 6 | 33 |
| 15 | Fabbri et al. (1980) | 2.9 | 0.95 | 256.7 | 13.8 | 57.4 | 7.7 | 300 |
| 16 | Boughn et al. (1981) | 3.78 | 0.30 | 275.4 | 3.9 | 46.8 | 4.5 | 46 |
| 17 | Cheng(1983) | 3.8 | 0.3 | | | | | 30 |
| 18 | Fixsen et al. (1983) | 3.18 | 0.17 | 265.7 | 3.0 | 47.3 | 1.5 | 25 |
| 19 | Lubin (1983) | 3.4 | 0.2 | | | | | 90 |
| 20 | Strukov et al. (1984) | 2.4 | 0.5 | | | | | 67 |
| 21 | Lubin et al. (1985) | 3.44 | 0.17 | 264.3 | 1.9 | 49.2 | 1.3 | 90 |
| 22 | Cottingham(1987) | 3.52 | 0.08 | 272.2 | 2.3 | 49.9 | 1.5 | 19 |
| 23 | Strukov et al. (1987) | 3.16 | 0.07 | 266.4 | 2.3 | 48.5 | 1.6 | 67 |
| 24 | Halpern et al. (1988) | 3.4 | 0.42 | 289.5 | 4.1 | 38.4 | 4.8 | 150 |
| 25 | Meyer et al. (1991) | | | 249.9 | 4.5 | 47.7 | 3.0 | 170 |
| 26 | Smoot et al. (1991) | 3.3 | 0.1 | 265 | 1 | 48 | 1 | 53 |
| 27 | Smoot et al. (1992) | 3.36 | 0.1 | 264.7 | 0.8 | 48.2 | 0.5 | 53 |
| 28 | Ganga et al. (1993) | | | 267.0 | 1.0 | 49.0 | 0.7 | 170 |
| 29 | Kogut et al. (1993) | 3.365 | 0.027 | 264.4 | 0.3 | 48.4 | 0.5 | 53 |
| 30 | Fixsen et al. (1994) | 3.347 | 0.008 | 265.6 | 0.75 | 48.3 | 0.5 | 300 |
| 31 | Bennett et al. (1994) | 3.363 | 0.024 | 264.4 | 0.2 | 48.1 | 0.4 | 53 |
| 32 | Bennett et al. (1996) | 3.353 | 0.024 | 264.26 | 0.33 | 48.22 | 0.13 | 53 |
| 33 | Fixsen et al. (1996) | 3.372 | 0.005 | 264.14 | 0.17 | 48.26 | 0.16 | 300 |
| 34 | Lineweaver et al. (1996) | 3.358 | 0.023 | 264.31 | 0.17 | 48.05 | 0.10 | 53 |

[a] Galactic coordinates

# References

1. Banday, A., et al. 1996, Ap.J., submitted
2. Bennett, C.L., et al. 1994, Ap.J., 436, 423
3. Bennett, C.L., et al. 1996, Ap.J., in press
4. Bersanelli, M., et al. 1996, A&A in press
5. Boughn, S.P. et al. 1971, Ap.J., 165, 439
6. Boughn, S.P. et al. 1981, Ap.J., 243, L113
7. Cheng, E.S. et al. 1979, Ap.J. 232, L139
8. Cheng, E.S. 1983 Ph.D. thesis, Princeton Univ.
9. Conklin, E.K. 1969, Nature, 222, 971
10. Conklin, E.K. 1972, IAU Symposium 44, ed. D.S. Evans (Dordrecht: Reidel), p 518
11. Corey, B.E. 1978, Ph.D. thesis Princeton U.
12. Corey, B.E. & Wilkinson D. T., 1976, Bull. Amer. Astron. Soc, 8, 351
13. Cottingham, D.A. 1987, Ph.D. thesis, Princeton Univ.
14. Fabbri, R., et al. 1980, PRL, 44, 1563, erratum 1980, PRL, 45, 401
15. Fixsen, D.J., Cheng, E.S. & Wilkinson, D.T. 1983, PRL, 50, 620
16. Fixsen, D.J., et al. 1994, Ap.J., 420, 445
17. Fixsen, D.J., et al. 1996, Ap.J., submitted
18. Ganga, K., Cheng, E., Meyer, S., Page, L. 1993, Ap.J., 410, L57
19. Gorenstein, M.V. 1978, Ph.D. thesis, U.C.Berkeley
20. Halpern, M., et al. 1988 Ap.J., 332, 596
21. Henry, P.S. 1971, Nature, 231, 516
22. Kogut, A., et al. 1993, Ap.J., 419, 1
23. Kogut, A., et al. 1996a, Ap.J.L., in press
24. Kogut, A., et al. 1996b, Ap.J., in press
25. Lineweaver, C.H., et al. 1995, Astrophysical Letters and Comm., 32, pp 173-181
26. Lineweaver, C.H., et al. 1996, Ap.J., in press
27. Lubin, P.M., Epstein, G.L., Smoot, G.F. 1983, PRL, 50, 616
28. Lubin, P.M. et al. 1985, Ap.J., 298, L1
29. Meyer, S.S., et al. 1991, Ap.J., 371, L7
30. Muehler, D. 1976, in Infrared and Submillimeter Ast., ed G.Fazio, D.Reidel, Dordrecht, p63
31. Partridge, R.B. & Wilkinson, D. T., 1967, PRL 18, 557
32. Penzias, A.A. & Wilson, R. W. 1965, Ap.J., 142, 419
33. Richards, P., (comment at this Moriond Meeting)
34. Smoot, G.F., Gorenstein, M. V. & Muller, R. A. 1977, PRL, 39, 898
35. Smoot, G.F. & Lubin, P. 1979, Ap.J. 234, L83
36. Smoot, G.F., et al. 1991, Ap.J., 371, L1
37. Smoot, G.F., et al. 1992, Ap.J., 396, L1
38. Strukov, I.A., Skulachev, D.P. 1984, Sov. Ast. Lett. 10, 3
39. Strukov, I.A., Skulachev, D.P., Boyarskii, M.N., Tkachev, A.N. 1987, Sov. Ast. Lett. 13, 2
40. Weinberg, S. 1972, Gravitation and Cosmology, (NY:Wiley), p 521
41. Wilkinson, D.T. & Partridge, R.B. 1969, Partridge quoted in American Scientist, 57, 37